\documentclass{article}
\begin{document}

\title{{\bf On the notion of conditional symmetry \\
of differential equations}}

\author{Giampaolo Cicogna\thanks{Email: cicogna@df.unipi.it}\quad  and\quad
Michele  Laino \\
\\
Dip. di Fisica ``E.Fermi'' dell'Universit\`a and I.N.F.N., Sez. di Pisa,
\\ Largo B. Pontecorvo 3, Ed. B-C, I-56127, Pisa, Italy}

\date{}

\maketitle

\begin{abstract}
Symmetry properties of PDE's are considered  within a systematic and unifying 
scheme: particular attention is devoted to the notion of conditional 
symmetry, leading to the distinction and a precise characterization 
of the notions of ``true'' and ``weak'' conditional symmetry.
Their relationship with exact and partial symmetries is also discussed. 
An extensive use of ``symmetry-adapted'' variables is made;
several clarifying examples, including the case of Boussinesq equation, 
are also provided.
\end{abstract}



\def \ov{\over}
\def \bar{\overline}
\def \beq{\begin{equation} }
\def \eeq{\end{equation} }
\def \lb{\label}
\def \nn{\nonumber}

\newtheorem{lemma}{Lemma}
\newtheorem{theorem}{Theorem}
\newtheorem{proposition}{Proposition}
\newtheorem{definition}{Definition}

\def \pd{\partial}
\def\~#1{\widetilde #1}
\def\.#1{\dot #1}
\def\^#1{\widehat #1}
\def\d{{\rm d}}       
\def \id{\! :=}
\def\dst{\displaystyle}

\def \sy {symmetry}
\def \sys {symmetries}
\def \so {solution}
\def \eq{equation}
\def \R{{\bf R}}

\def\a{\alpha}
\def\be{\beta}
\def\phi{\varphi}
\def\de{\delta}
\def\g{\gamma}
\def\De{\Delta}
\def\th{\theta}
\def\ka{\kappa}
\def\s{\sigma}

\def \qq{\qquad}
\def \q{\quad}
\def \pn{\paragraph\noindent}
\def \sk{\medskip}
\def \ni{\noindent}
\def\BE{Boussinesq equation}
\def\CS{conditional symmetry}
\def\={\, =\, }

\section{Introduction}

In the study of general aspects of differential \eq s, and also in the
concrete problem of finding  their explicit \so s, a fundamental role is 
played, as well known, 
by the analysis of \sy\ properties of the \eq s. In addition to the
classical notion of Lie ``exact'' \sys\ (see e.g. \cite{Ov}--\cite{BA}), 
an important class of \sys\ is given by the ``conditional \sys'' (or 
``nonclassical \sys''), introduced and developed by Bluman and Cole 
\cite{BC,BC1}, Levi and Winternitz \cite{LW,W2}, Fushchych \cite{FK,Fu} 
and many others (see e.g. \cite{Ib,W2}).

In this paper we will be concerned with partial differential \eq s (PDE) 
and with the above mentioned types of \sys , and 
also with the   notion of ``partial \sy '', as defined in
\cite{CG}: in the context of a simple comprehensive scheme,  
we will distinguish different notions of conditional \sy , with
a precise characterization of their properties and a clear comparison 
with other types of \sys .

An extensive use will be made of the ``symmetry-adapted'' variables 
(also called ``canonical coordinates'', see e.g.  \cite{Ol,BA} and also 
\cite{CiK}), which 
reveal to be extremely useful; several  clarifying examples will be 
also provided, including the case of \BE , which offers good examples 
for all the different notions of \sy\ considered in this paper (see also
\cite{Cl}). 

For the sake of simplicity, only ``geometrical'' or Lie point-\sys\  
will be considered, although the relevant results could be extended
to more general classes of \sys , as 
generalized or  B\"acklund,  potential or nonlocal \sys , whose 
importance is well known and also recently further emphasized (see e.g. 
\cite{GMR}--\cite{Sop}).


\section{Preliminary statements}

Let us start with a preliminary Lemma,  simple but  
important for our applications. In view of this, the notations 
are chosen similar, as far as possible, to those used below.

\begin{lemma} Consider a system of $n$ \eq s for the $n$ functions
$y_a=y_a(s)$ ($a=1,\ldots,n$; $s\in\R$) of the form (sum over repeated
indices)
\beq
{\d y_a\over{\d s}}\= G_{ab}(s,y)\, y_b \lb{ygy}
\eeq
where $G_{ab}$ are $n\times n$ given functions of $s$
and of $y\equiv(y_1,\ldots,y_n)$, which are assumed regular enough
(e.g., analytic in a neighbourhood of $s=0,\,y=0$). Then, any \so\ of
(\ref{ygy}) can be written, in a neighbourhood of $s=0$,  
\beq
y_a(s)\=R_{ab}(s)\, \ka_b \lb{yGk}
\eeq
where $\ka\equiv(\ka_1,\ldots,\ka_n)$ are constants, and $R_{ab}$ are regular
functions with
\[ R_{ab}(0)\= \de_{ab} \]
(then, $\ka_a=y_a(0)$). Reciprocally, for any \so\
$y_a(s)$, there are regular functions $S_{ab}(s)$ such that
\beq \ka_a\= S_{ab}(s)\, y_b(s) \ . \lb{kSy} \eeq
\end{lemma}
{\it Proof.} The result is nearly trivial if $G_{ab}$ do not depend
on $y$. In the general case, let $y_a=\bar{y}_a(s)$ denote any given \so\ of
(\ref{ygy}) in a neighbourhood of $s=0$, determined by $n$ initial values 
$\ka_a=\bar{y}_a(0)$ (we omit to write explicitly the dependence on the
$\ka$). Let us put
\[ K_a(s)\= S_{ab}(s)\,\bar{y}_b(s)  \]
where $S_{ab}$ are functions to be determined; we then get (with $'=\d/\d s$)
\beq K'_a\=S'_{ab}\,\bar{y}_b+S_{ab}\,\bar{y}'_b\=
\Big(S'_{ab}+S_{ac}\,G_{cb}(s,\bar{y}(s))\Big)\,
\bar{y}_b \lb{ssg} \eeq
Consider now the \eq\ for the matrix $S$, with clear notations in matrix form,
\beq S'\=-SG \lb{SSG} \eeq
where it is understood that the generic \so\ $\bar{y}$ is replaced in $G$ by its 
expression depending on $s$ (and on $\ka$, of course). 
Now, eq. (\ref{SSG}) always admits a \so\ $S$ --~as well known
(see e.g. \cite{CL}) --
which can be characterized as a fundamental matrix for the associated
``adjoint'' system $\zeta'_a=-\zeta_b G_{ba}$. In particular, this
fundamental matrix can be constructed assuming as initial
value at $s=0$ the matrix $S(0)=I$.   
Therefore, choosing this matrix $S$, one gets from (\ref{ssg})
$K_a= $ const = $K_a(0)=S_{ab}(0)\bar{y}_b(0)=\bar{y}_a(0)=\ka_a$, and
$\ka_a=S_{ab}(s)\,\bar{y}_b(s)$. The matrix
$S(s)$ can be locally inverted, giving  for any \so\
$y_a(s)$, $y_a=S^{-1}_{ab}\, \ka_b\id R_{ab}\,\ka_b$ with
$R=R(s)$ and $R(0)=I$.

\medskip

\ni
{\it Remark 1}. As is clear from the proof, $S$ and $R$ also depend
on the initial values $\ka$ which indeed determine the generic \so\ $\bar{y}(s)$;
the only relevant points here are the ``factorization'' of the $\ka$ as 
in (\ref{yGk}) and the form (\ref{kSy}), i.e. the possibility of obtaining   
$s-$independent ``combinations'' (with coefficients $S$ depending 
on $s$) of the components of each \so .
In our applications, the functions $G_{ab}$ will also
depend on some other parameters; then all results hold true, but
clearly $S,\, R$ and
$\ka$ turn out to be functions of these additional quantities.

\sk

In the following, we will consider systems of PDE's, denoted by
\begin{eqnarray} \De\equiv \De_a(x,u^{(m)})\=0 \q , &
\q a=
,\ldots,\nu\, , \\ \q u\equiv(u_1,\ldots,u_q) \q ; &
\q x\equiv(x_1,\ldots,x_p) \nonumber \lb{De1} \end{eqnarray}
for the $q$ functions $u_\a=u_\a(x)$ of the $p$ variables $x_i$,
where $u^{(m)}$ denotes the functions $u_\a$ together with their
$x$ derivatives up  to the order $m$,
with usual notations and assumptions (as stated, e.g., in \cite{Ol}).
In particular, we will always assume that all standard smoothness
properties and the maximal rank condition are satisfied.
As anticipated, only Lie point-\sys\ will be considered, with
infinitesimal generator given by vector field
\beq X\= \xi_i(x,u){\pd\over {\pd x_i}}+\phi_\a(x,u){\pd \over{\pd u_\a}}
\lb{X1}
\eeq
To simplify notations, we shall denote by $X^*$ the ``appropriate''
prolongation of $X$ for the \eq\ at hand, or -- alternatively -- its
infinite prolongation (indeed, only a finite number of terms will appear
in calculations).

For completeness, and -- even more -- for comparison with the subsequent
Definition 2, let us start with the following (completely standard)
definition (cf. \cite{Ol}).

\begin{definition}
A (nondegenerate) system of  PDE $\De_a(x,u^{(m)})\=0$
is said to admit the Lie point-\sy\ generated by the vector field $X$
(or to be symmetric under $X$) if the following condition
\beq X^*(\De)|_{\De=0}\=0 \lb{XXDD} \eeq
is satisfied, or -- equivalently (at least under mild hypothesis, see 
\cite{Ol}) -- if there are functions $G_{ab}(x,u^{(m)})$ such that
\beq (X^*(\De))_a\=G_{ab}\,\De_b \ .\lb{XDgD}\eeq
\end{definition}

\sk
\ni
Let us also give this other definition.
\begin{definition}
A system of PDE as before is said to be {\rm invariant} under 
a vector field  $X$  if
\beq X^*(\De)\=0 \ . \lb{inv}\eeq
\end{definition}

\sk
For instance, the Laplace \eq\ $u_{xx}+u_{yy}=0$ is {\em invariant}
under the
rotation \sy\ generated by $X=y\pd/\pd x-x\pd/\pd y$; the heat \eq\
$u_t=u_{xx}$ is {\em symmetric} but {\em not} invariant under
\beq X=2t{\pd\ov \pd x}-xu{\pd\ov \pd u} \lb{Xhea} \eeq
indeed one has
$\, X^*(u_t-u_{xx})= -x (u_t-u_{xx})$.

\def\"#1{\ddot #1}

\section{Symmetric and Invariant Equations}

Let us introduce a first simplification:  we will assume that
the vector fields $X$ are ``projectable'', or -- more explicitly --
that
the functions $\xi$ in (\ref{X1}) do not depend on $u$, as often happens
in the study of PDE's. This strongly simplifies 
calculations, especially in the introduction of the more ``convenient'' or
``symmetry-adapted'' variables, and allows a more direct relationship
between  \sys\ and  symmetry-invariant \so s, as discussed in  
\cite{Pu}.

A first result, concerning ``exact'' (to be distinguished from
conditional or partial, see below) \sys\ is the following (see also 
\cite{Ol,BA}).

\begin{theorem}
Let $\De=0$ be  a nondegenerate system of PDE, symmetric under a
vector field $X$, according to Def.~1. Then, there are  new $p+q$
variables
$s,z$ and $v$,  with $s\in\R$, $z\in\R^{p-1}$,
$v\equiv(v_1(s,z),\ldots,v_q(s,z))$, and a new system of PDE's, say
$K=0$, with $\q K_a\=S_{ab}(s,z,v^{(m)})\,\~\De_b(s,z,v^{(m)})$ [where
$v^{(m)}$  stands for $v(s,z)$ and its derivatives with respect to 
$s$ and $z$, and $\~\De=\~\De(s,z,v^{(m)})$
is $\De$ when expressed in terms of the new variables $s,z,v$],
which is  locally {\rm equivalent} to the initial system and is {\rm
invariant} (as in Def.~2) under the \sy\ $X=\pd/\pd s$, i.e.
$K_a=K_a(z,v^{(m)})$.
\end{theorem}
{\it Proof.} Given  $X$, one introduces ``canonical variables'' $s,z$,
defined by
\beq
X\ s\,\equiv\,\xi_i {\pd s\over{\pd x_i}}+\phi_\a {\pd s\over{\pd
u_\a}}\=1 \q ; \q X\ z_k \= 0 \q (k=1,\ldots, p-1)
\lb{Xsz}\eeq
One first considers the subset of characteristic \eq s $\d x_i/\xi_i=\d s$
which do not contain the variables $u_\a$, and finds the variable $s$
together with the $X-$invariant
variables $z_k$. Then, using the characteristic \eq s $\d
x_i/\xi_i=\d u_\a/\phi_\a$, one finds the $q$ invariant quantities
$v_\a$, and expresses the $u_\a$ in terms of $v_\a$ and of
the new independent variables $s,z_k$. Once written in these
coordinates, the \sy\ field and all its prolongations are simply given
by
\beq X=X^*={\pd\ov \pd s} \lb{XX*}\eeq
whereas the \sy\ condition becomes 
$\pd\~\De/\pd s|_{\~\De=0}\= 0$ or
\beq 
{\pd\over{\pd s}}\~\De_a\=G_{ab}\~\De_b
\lb{dsDgD} \eeq
An application of Lemma 2.1, where the
role of $y$ is played here by $\~\De$ and that of $\ka$ by $K$, shows
that there are suitable ``combinations'' $K_a=S_{ab}\~\De_b$ of the
$\~\De_a$ which do not depend {\em explicitly} on $s$, i.e.
$(\pd/\pd s)K=0$.

\medskip
This result can be compared with an analogous result presented in
\cite{CDW}, where however the point of view is different (i.e.,
constructing \eq s with a prescribed algebra of \sys ).

\medskip
\ni
{\em Example 1}. Consider the quite trivial system of PDE for $u=u(x,y)$
$$  u_{xx}+u_{yy}+u_{xxx}\= 0 \eqno(15a)$$
$$  u_{xxx}\=u_{xxy}\=u_{xyy}\=u_{yyy}\=0 \eqno(15b) 
\addtocounter{equation}{1}$$
This system admits the rotation \sy\ $X=y\pd/\pd x-x\pd/\pd y$, although
none of the \eq s above is invariant or symmetric under rotations. The
variables $s,z$ are in this case obviously the polar variables $\theta,
r$, and  $X=X^*=\pd/\pd \th$; it is now easy to construct
combinations of the above \eq s for $v=v(r,\th)$ which are invariant
under $\pd/\pd\th$: e.g.
\[ yu_{xxx}-
           x u_{xxy} + y u_{xyy} -
              x u_{yyy} =
         - r^{-2}(r v_{r \th} + r^2 v_{rr \th} +
                   v_{\th \th \th}) = 0
\]
\[ xu_{xxx}
          + y u_{xxy} + x u_{xyy} +
             y u_{yyy} = (r^{-2})
         (-r v_r + r^2 v_{r r} - 2
v_{\th \th} + r^3 v_{rrr} + r  v_{r \th \th}) = 0 \]

It can be remarked that considering   \eq\ (15$a$), together with
only the first one of the (15$b$), i.e. $u_{xxx}=0$, one obtains a
system which is {\em not} symmetric under rotations, although the \eq\
$u_{xxx}=0$ expresses the vanishing of the ``\sy\ breaking term'' in
(15$a$). As a  consequence, the system of these two \eq s
would admit \so s, e.g. $u=x^2y-y^3/3$, which are {\it not} transformed by 
rotations into other \so s. 
$\hfill\odot$

\sk
\ni
{\em Example 2}.
In the example of heat \eq\ mentioned at the end of previous Section, 
choosing the variables
$s=x/2t,\, z=t$ and with $u=\exp(-zs^2)v(s,z)$ as determined by the \sy\
vector field (\ref{Xhea}), the \eq\ is transformed into the {\em equivalent}
\eq\ for $v=~v(s,z)$
\[ 4z^2v_z+2zv-v_{ss}\=0 \]
($v_s=\pd v/\pd s$, etc.)
which indeed does not depend explicitly on $s$ and therefore is
{\em invariant} under the \sy\ $X=\pd/\pd s$ (but does contain a function
$v$ depending on $s$). Now looking for  \so s with $v_s=0$, i.e.
with $v=w(z)$, one obtains the known reduced \eq\ 
$2zw_z+w\= 0$ (see \cite{Ol}). $\hfill\odot$

\medskip
It should be emphasized that the result in Theorem 1 is
 not the same as (but is related to, and includes in particular) the well 
known result concerning
the reduction of the given PDE to $X-$invariant \eq s for the 
variables $w(z)$: indeed, introducing the new ``\sy-adapted'' variables 
$s,z$ and $v(s,z)$, we have transformed the \eq\ into a locally 
{\em equivalent} \eq\ for $v(s,z)$. If one now {\em further} assumes
that $\pd v/\pd s=0$, i.e. if one looks for the $X-$invariant \so s
$v=w(z)$, then the \eq s $K_a=0$ become a system of \eq s
\beq K_a^{(0)}(z,w^{(m)})=0 \lb{K0}\eeq
involving only the variables $z$ and functions depending only on $z$
(see \cite{ZTP} for a detailed discussion on the reduction procedure).
In particular, in the case of a single PDE for a single unknown function 
depending on two variables, the PDE is reduced to an ODE, as well known,
and as in Example 2 above.

\section{Conditional Symmetries, in ``true'' and ``weak'' sense}

The above approach includes in a completely natural way some other
important situations. It is known indeed that, by means of the
introduction of the notion of conditional \sy\ (CS), one may
obtain other \so s which turn out to be invariant under these
``nonclassical'' \sys\ \cite{Ib}, \cite{BC}--\cite{Fu}. But there are  
{\em different} types of CS,  and it is useful to distinguish these 
different notions and to see how they can be fitted in this scheme.

To avoid unessential complications with notations, we will consider
from now on only the case of a single PDE $\De=0$ for a single unknown 
function $u(x)$; the extension to the general cases is in principle 
straightforward.

As well known, a vector field $X$ is said to be a \CS\ for the
\eq\ $\De=0$ if $X$ is an ``exact" Lie point-symmetry for the system
\beq \De\, =\, 0 \qq ; \qq X_Q u\equiv \phi-\xi_i {\pd u\ov \pd
{x_i}}\=  0 \lb{gcs} \eeq
where $X_Q$ is the \sy\ written in ``evolutionary form'' \cite{Ol}.
The second
\eq\ in  (\ref{gcs}), indicating that we are  looking 
for \so s  {\it invariant} under $X$,
is automatically symmetric under $X$; we
have then only to impose 
\beq X^* (\De)\Big|_\Sigma\, =\, 0 \lb{gc*} \eeq
where $\Sigma$ is the set of the simultaneous \so s of the two \eq s
(\ref{gcs}),  plus (possibly) some differential consequences of 
the second one 
(see \cite{Ol}, \cite{PS}--\cite{Pop} for a precise and detailed
discussion on this  point and the related notion of degenerate systems of
PDE). In the canonical variables $s,z$ and $v=~v(s,z)$ 
determined by the vector field $X$,
the invariance condition $X_Qv=0$ becomes
\beq {\pd v\ov \pd s}\=0 \lb{us}\eeq
and the condition of CS (\ref{gc*}) takes the 
simple form (let us now retain for simplicity the same notation $\De$, 
instead of $\~\De$, also in the new coordinates)
\beq {\pd \De\over {\pd s}}\Big|_{\Sigma}\=0 \lb{DsS}\eeq
here $\Sigma$ stands for the set of the simultaneous \so s of $\De=0$
and   $v_s=~0$, together with the derivatives of $v_s$
with respect to all variables $s$ and $z_k$. Using  the
global notation $v_s^{(\ell)}$ to indicate $v_s,v_{ss},v_{sz_k}$ etc.,
the CS condition (\ref{DsS}) becomes then equivalently, according to Def.
1, and with clear notations, 
\beq {\pd\ov \pd
s}\De\=G(s,z,v^{(m)})\,\De+\sum_\ell H_{\ell}(s,z,v^{(m)})\,v_s^{(\ell)}
\lb{dgh}\eeq
which, in the original coordinates $x,u$, states that $X^*\De$ is a 
``combination'' of $\De$ and of $X_Qu$ with its differential consequences. 
Now, another application of Lemma~2.1 (the role of $y$
being played by $\De$ and $v_s^{(\ell)}$) gives that
$\De$ must have the form
\beq \De\=R(s,z,v^{(m)}) K(z,v^{(m)})+\sum_\ell\Theta_\ell(s,z,v^{(m)})\,
v_s^{(\ell)} \lb{DRT}\eeq
where the points to be emphasized are that $R,K$ do not contain
$v_s^{(\ell)}$ and that $K$ does not depend {\em explicitly} on $s$.

If one now looks for \so s of $\De=0$ which are independent on $s$, i.e.
$v=w(z)$ and $v_s^{(\ell)}=0$, 
then eq. (\ref{DRT}) becomes a ``reduced'' \eq\ $K^{(0)}(z,w^{(m)})=0$,
just as in the exact \sy\ case.

\sk
\ni
{\em Remark 2}. If $X$ is a CS for a differential \eq , then clearly also
$X_\psi=\psi(x,u)X$, for any smooth function $\psi$, is another CS. While the
invariant variables $z$ are the same for $X$ and $X_\psi$, the variable
$s$ turns out to be different. This implies that, writing the differential 
\eq\ in terms of the canonical variables, one obtains in general different 
\eq s for different choices of $\psi$. All these \eq s will produce the same
reduced \eq\ when one looks for invariant \so s $v=w(z)$. 

\sk\ni
{\em Example 3}. It is known that the nonlinear acoustic \eq\
\cite{FK,Fu,OR}  
\[ u_{tt}\=u\,u_{xx}\qq ; \qq u=u(x,t) \]
admits the CS
\[ X\=2t{\pd\ov \pd x}+{\pd\ov \pd t}+8t{\pd\ov \pd u} \]
Introducing the canonical variables $s=t,z=x-t^2$ and $u=v(s,z)+4s^2$,
the \eq\ becomes
\beq   8 - 2  v_z - v  v_{z z} + v_{s s} - 4  s  v_{s z}  \= 0
\lb{aeq1} \eeq
Considering instead $(1/2t)X=\pd/\pd x+(1/2t)\pd/\pd t+4\pd/\pd u$,
one gets $s=~x$, $z=~x-t^2, u=v(s,z)+4(s-z)$ and the \eq\
becomes
\beq    8 - 2 v_z - v v_{z z} - 4 s v_{s s} +
               4 z v_{s s} - 2 v v_{s z} - 8 s v_{s z} +
               8 z v_{s z} - v v_{ss} \= 0 \lb{aeq2}\eeq
Both \eq s (\ref{aeq1}) and (\ref{aeq2}) have the form (\ref{DRT}), as
expected, and both become the reduced ODE 
$\, 8-2w_z-ww_{zz}\=0$.
 $\hfill\odot$

\sk
The presence of some terms containing $s$ in the above \eq s
(\ref{aeq1},\ref{aeq2})  shows that $X$ is not an exact \sy , and the fact
that these  terms disappear when $v_s=0$ shows that $X$ is a CS.  

However,  the above one is not the only way to obtain reduced \eq s.

Indeed, the rather disappointing remark is that, as pointed out
by Olver and Rosenau \cite{OR} (see also \cite{PS}),
given an {\em arbitrary} vector field $X$, if one  can find
some particular simultaneous \so\ $\^u$ of the two \eq s (\ref{gcs}),
then the CS condition (\ref{gc*}) turns out to be  automatically 
satisfied when  evaluated along this \so , i.e.:
\beq X^*(\De)|_{\^u}\=0 \ . \lb{*^}\eeq
It can be interesting to verify this fact in terms of the canonical
variables $s,z,v$: indeed one has ($\d/\d s$ is the total derivative)
\beq X^*(\De)\={\pd \De\ov \pd s}\={\d\De\ov \d s}-v_s^{(\ell)}{\pd\De\ov
\pd v^{(\ell)}} \lb{dds}\eeq
which vanishes if one chooses a \so\ of $\De=0$ of the form $\^v=\^w(z)$. 
Even more, it is enough to find an {\em arbitrary} \so\ of $\De=0$; then,
choosing any vector field  leaving invariant this \so , one could
conclude that -- essentially~-- any vector field
is a CS, and any \so\ is invariant under some CS: cf. \cite{OR}~! 
This issue has been also considered  in \cite{Odm}, from another point of 
view (see also the end of this section).

The point is that the existence of   some  \so\ $\^u$ of the
two \eq s (\ref{gcs})
is  not exactly equivalent to the condition (\ref{gc*}), this 
happens essentially because $X$ in this case is a \sy\ of an {\it 
enlarged} system which includes the compatibility conditions of the 
differential consequences of both \eq s in (\ref{gcs}) (or the 
``integrability conditions'': see \cite{Ol},\cite{PS}--\cite{On}). Therefore 
it is important to clearly distinguish {\em different} notions 
and introduce a sort of ``classification'' of CS.

We will say that $X$ is CS in ``true'' or standard sense if 
$X^*(\De)|_\Sigma=~0$ is
satisfied: the discussion and the examples above cover precisely this
case;  also the examples of CS considered in the literature are usually
CS of  standard type (see e.g. \cite{BC}-\cite{Fu},
\cite{PS1}-\cite{Pop}, but see also
\cite{Zh,OR,Odm,FV}). Instead, when $X^*(\De)|_{\^u}=~0$ is satisfied
only for some
$\^u$, we  shall say that a ``weak'' CS is concerned (we will be more
precise in a  moment;  notice  however that some authors 
call generically ``weak'' \sys\ all non-exact \sys ). What happens in 
this case is -- once again~-- more clearly seen in the canonical
variables  determined by the given vector field $X$ (see also \cite{CiK}):
assume indeed   for a moment that in these coordinates the PDE takes
the form
\beq \lb{sr} \De\=\,\sum_{r=1}^\s s^{r-1} K_r(z,v^{(m)})+
\sum_\ell\Theta_\ell(s,z,v^{(m)})\, v_s^{(\ell)} \=0 \eeq
where the part not containing $v_s^{(\ell)}$ is a polynomial in the 
variable $s$, with coefficients $K_r$ not depending explicitly on $s$.
Now, if one looks for $X-$invariant \so s $w(z)$ of $\De=0$, one no 
longer obtains reduced \eq s involving only the invariant
variables  $z$ and $w(z)$, as in the case of Eq. (\ref{DRT}), but one 
is faced (cf. \cite{OR,On})  with the system of reduced \eq s (not
containing $s$ nor functions of $s$)
\beq K_r^{(0)}(z,w^{(m)})\=0\qq ; \qq r=1,\ldots,\s \lb{Krz}\eeq
Assume that this system admits some \so\ (it is known that the 
existence of invariant \so s is by no means
guaranteed in general, neither for ``true'' CS, nor
for ``exact'' Lie \sys ), and denote by $\Sigma_\s$ the set of these \so s: 
for any $\^w(z)\in\Sigma_\s$ we are precisely in the case of weak CS.

The identical conclusion holds if the initial PDE is transformed into an 
expression of this completely general form (instead of (\ref{sr}))
\beq \De\=\sum_{r=1}^\s R_r(s,z,v^{(m)})\, K_r(z,v^{(m)})+
\sum_\ell\Theta_\ell(s,z,v^{(m)})\, v_s^{(\ell)}\=0  \lb{Kr}\eeq
with the presence here of a sum of $\s$ terms $R_rK_r$ (with $\s >1$), where the 
coefficients $R_r$ which depend on $s$ are grouped together, with the only 
obvious condition that the coefficients $R_r$ be linearly independent (the idea 
should be  that of obtaining the minimum  number of independent 
conditions (\ref{Krz})).

We now see that the set $\Sigma_\s$  can be characterized equivalently 
as the set of the \so s of the system
\beq \De\=0 \q ;\q {\pd \De\ov \pd s}\=0\q ;\ldots ; \q
{\pd^{\s-1}\De\ov \pd s^{\s-1}}\=0 ;\q v_s^{(\ell)}\=0
\lb{d1s}\eeq
(indeed   the $R_r$ are also functionally independent as 
functions of~$s$). 

Conversely, if a $\De(s,z,v^{(m)})=0$ is such that a system like 
(\ref{d1s}) admits the \sy\ $X=\pd/\pd s$, then condition (\ref{XDgD}) 
must be satisfied, and applying once again Lemma 2.1, we see that $\De$ 
must have the form (\ref{Kr}).
 
Therefore (\ref{Kr}) is the most general form of an \eq\ exhibiting 
the {\em weak} CS $X=\pd/\pd s$, to be compared with (\ref{DRT}), 
which corresponds to the case of {\em true} CS.

\medskip
Let us now come back to the original coordinates $x,u$: we will see that 
the set of conditions (\ref{d1s}) is the result of the following procedure.

Given the \eq\ $\De=0$, and  a vector field $X$, assume   that the
system of \eq s (\ref{gcs}) is {\em not} symmetric under $X$ (therefore,
that $X$ is not a ``true'' CS for $\De=0$), then put
\beq \De^{(1)}\id X^*(\De) \lb{D1}\eeq
and consider $\De^{(1)}=0$ as a new condition to be fulfilled, obtaining 
in this way the augmented system (the first step of this approach is similar to a procedure, 
involving contact vector fields, which has been proposed in \cite{DV})
\beq \De\=0 \qq ; \qq \De^{(1)}\=0 \qq ;\qq X_Qu\=0 \lb{dd1}\eeq
If this system is symmetric under $X$, i.e. if
\beq X^*(\De)|_{\Sigma_1}\=0 \lb{xs1}\eeq
where $\Sigma_1$ is the set of simultaneous \so s of (\ref{dd1}), we
can say that $X$ is ``weak CS of order 2'' (according to this, a true CS
is of order 1). If instead (\ref{dd1}) is not symmetric
under $X$, the procedure can be iterated, introducing
\beq \De^{(2)}\id X^*(\De^{(1)}) \lb{dd2}\eeq
and appending the new \eq\ $\De^{(2)}=0$, 
and so on. Finally, we  will say that $X$ is a weak CS of order $\sigma$ if
\beq X^*(\De)|_{\Sigma_\sigma}\=0 \lb{xss}\eeq
where $\Sigma_\sigma$ is the set (if not empty, of course) of the
\so s of the system
\beq \De\=\De^{(1)}\=\ldots\=\De^{(\s -1)}\=0\q ,\q X_Q\,u\=0 
\lb{dddd}\eeq
(as already pointed out, it is understood -- here and in the following -- 
that also the differential consequences of $X_Qu=0$ 
must be taken into account; clearly, the additional conditions $\pd 
\De/\pd s=0$ or $X^*\De=0$ and so on, should not be confused with the 
differential consequences of the \eq\ $\De=0$).

\sk\ni
{\em Remark 3}  (The ``partial'' \sys ). The above procedure for finding 
weak CS is reminiscent of 
the procedure used for constructing {\em partial symmetries}, according to the 
definition proposed in \cite{CG} (see also \cite{C}),  the  (relevant !)
difference being the presence in the weak CS case  of the additional
condition
$X_Qu=0$. Let us recall indeed that a vector field $X$ is said to be a
partial \sy\ of order $\s$ for $\De=0$ if the set of \eq s, with the
above definitions (\ref{D1},\ref{dd2}), 
\beq \lb{psa} \De\=\De^{(1)}\=\ldots\=\De^{(\s -1)}\=0 \eeq
admits some \so s. In terms of the variables $s,z$ and $v(s,z)$,
conditions (\ref{psa}) are the same as (\ref{d1s}) but {\em
without} the conditions 
$v_s^{(\ell)}=0$. The set of  \so s found in the presence of a partial
\sy\ provides   a ``symmetric set of \so s'', meaning that the \sy\
transforms a \so\ belonging to this set into a -- generally different --
\so\  in the same set.  If, in particular, this set includes some \so s 
which are left fixed by $X$, then this \sy\ is also a CS, either true
or weak. So, we could call the weak CS, by analogy, ``partial conditional
\sys\ of order
$\s$''. 

\sk
We can now summarize our discussion in the following way.
\begin{definition}
Given a PDE $\De=0$, a projectable vector field $X$ is a ``true'' 
conditional \sy\
for the \eq\ if it is  a \sy\ for the system
\beq   \De\=0\q ;\q  X_Qu\=0 \ . \lb{tcs}\eeq
A vector field $X$ is a ``weak'' CS (of order $\s$) if it is a \sy\ of the system
\beq \lb{wcs} 
\begin{array}{l} 
\De\=0 \  ;\  \De^{(1)}:=X^*(\De)\=0\ ;\   
\De^{(2)} := X^*(\De^{(1)})\=0\  ;\ \ldots ;\ \De^{(\s-1)}=0\   ;\\
~\\
X_Qu \= 0  \ .
\end{array}\eeq
\end{definition}

\sk

\begin{proposition}
If $X$ is a true CS, the system (\ref{tcs}) gives rise to 
a reduced \eq\ in $p-1$ independent variables, which -- if admits \so s --
produces   $X-$invariant \so s of  $\De=0$. If $X$ is a weak CS of order $\s$, 
the system (\ref{wcs}) gives  rise to  a system of $\s$
reduced \eq s, which --  if admits \so s -- produces  $X-$invariant  \so s of
$\De=0$.  Introducing $X-$adapted variables $s,z$, such that $Xs=1,\,
Xz=0$, the  PDE has the form (\ref{DRT}) in the case of true CS, or
(\ref{Kr}) in the case  of weak CS.
\end{proposition}

We can then rephrase the Olver-Rosenau statement \cite{OR} saying that 
{\em any vector field $X$ is either an exact, or a true CS, or a weak CS}. 
Similarly, rewriting the \eq\ $\De=0$ as in (\ref{sr}) or in (\ref{Kr})
but {\em without isolating} the terms $v_s^{(\ell)}$, we can also say, 
recalling the procedure used for obtaining partial \sys , that {\em any $X$ 
is either an exact or a partial \sy }. It is clear however that, as
already remarked, the set of  \so s which can
be  obtained in this way may be empty, or 
contain only trivial \so s (e.g., $u=$ const).

\sk\ni
{\em Example 4}. It is  known  that Korteweg-de Vries \eq
\[  \De\id u_t+u_{xxx}+uu_x\=0 \qq\qq u\=u(x,t) \]
does not admit (true) CS (apart from its exact \sys ). There are however 
weak CS; e.g. it is simple to verify that the scaling vector field
\[ X\=2x{\pd\ov{\pd x}}+t{\pd\ov{\pd t}}+u{\pd\ov {\pd u}} \]
is indeed an exact \sy\ for the system $\De=0,\,\De^{(1)}=X^*(\De)=0$ and 
$X_Qu=0$, and therefore is a weak CS of order $\s=2$, and   $u=x/t$ is 
a scaling-invariant \so . But also, if we consider only the system 
$\De~=~0,\,\De^{(1)}=X^*(\De)=0$ (i.e. without the invariance condition 
$X_Qu=0$), we obtain the symmetric set of \so s 
\[ u={x+c_1\ov{t+c_2}} \qq (c_1,c_2\={\rm const})\]
showing that $X$ is also a partial \sy . $\hfill\odot$

\sk

Few words, for completeness, about  the so-called ``direct method'' 
\cite{PS,Odm},
\cite{CK}--\cite{Nu} for finding \so s to PDE's.
The simplest and typical application of this method deals with PDE
involving a function of two
variables (which we shall call $x,t$, in view of the next applications), 
and one looks for \so s of the form
(also called ``similarity reduction \so s'')
\beq\lb{uwi} u(x,t)\=U(x,t,w(z)) \q{\rm with}\q z=z(x,t)\eeq
or -- more simply -- of the form (according to a
remark by Clarkson and  Kruskal \cite{CK}, this is
not a restriction, see also Lou \cite{Lo})
\beq \lb{uwz} u(x,t)\=\a(x,t)+\be(x,t)\, w(z) \q {\rm with} \q z=z(x,t)\eeq
one then substitutes (\ref{uwz}) into the PDE and {\em imposes} that
$w(z)$ satisfies an ODE.
Although this method is not based on any \sy , there is clearly a close
and   fully investigated relationship with \sy\ properties; referring to 
\cite{Pu,ZTP,Nu} for a complete and detailed discussion, we only
add here the following remark, to illustrate the idea in the present
setting. Assuming in (\ref{uwz}) that $z_t\not= 0$, one can choose 
$x$ and $z$ as independent variables, and then write (\ref{uwz}) in the 
form $u=\~u(x,z)=\~\a(x,z)+\~\be(x,z)w(z)$. Then, putting
\beq X\=\xi(x,z){\pd\ov \pd x}+\zeta(x,z){\pd\ov \pd z}+
\phi(x,z,u){\pd\ov \pd u} \lb{Xtt}\eeq
one can fix $\zeta=0$, in such a way that $X\, z=0$, choose $\xi=1$,
and finally impose that $X_Qu\equiv \~\a_x+\~\be_xw(z)-\phi(x,z,u)=0$
in order to determine the coefficient $\phi$ in (\ref{Xtt}). Then, by
construction, $\~u(x,z)$ is invariant under this $X$. It is also
easy to see that the invariance condition $X_Qu=0$ is satisfied exactly
by the family (\ref{uwz}). If $z_t=0$ in (\ref{uwz}), the same result is
true retaining $z=x$ and $t$ as independent variables, and choosing
\beq X\={\pd\ov\pd t}+ \phi(x,t,u){\pd\ov \pd u} \lb{Xtp}\eeq
So, the direct method has produced a set of \so s to the given PDE
which also satisfies the invariance condition $X_Qu=0$; then,
according to our discussion, $X$ is a CS for the PDE: it is a  
true CS if $w(z)$ satisfies a single ODE, as is usually the case in
the direct method, or a   weak  CS if the method has produced a
separation of the PDE into a system of ODE's.

Notice that a generalization of this method has been proposed in 
\cite{Gal}, with the introduction of {\em two} functions of the 
similarity variable $z$; this procedure has been further extended in 
\cite{Odm}, where its relationship with method of differential 
constraints is also carefully examined.

Other reduction procedures, based on the introduction of 
suitable multiple  differential constraints, have been also proposed, aimed 
at finding nonclassical \sys\ and \so s of differential problems:
see, e.g., \cite{GMR},\cite{Vo}-\cite{Kap}, and also \cite{Ib}.
It can be also remarked that in our 
discussion  we have only considered the case of a single vector 
field $X$; clearly, the situation becomes richer and richer if more than 
one vector field is taken into consideration. First of all, the reduction 
procedure itself must be adapted and refined when the given \eq\ admits 
an algebra of \sys\ of dimension larger than $1$ (possibly infinite): for 
a recent discussion see \cite{GTW}.

\section{Examples from the Boussinesq equation}

The \sy\ properties of the \BE\ 
\beq \lb{BE}
\De\id u_{tt}+u_{xxxx}+u\, u_{xx}+u_x^2\= 0\qq ; \qq u=u(x,t)
\eeq
have been the object of several papers (see e.g. \cite{LW,Cl,GMR}), but
it is  useful to consider here some special cases to illustrate the above 
discussion.
First of all, let us give the {\em invariant} form (according to Theorem
1) of the \eq\ under the (exact) dilational \sy
\beq \lb{D} D\=x{\pd\over{\pd x}}+2t{\pd\over{\pd t}}-2u
{\pd\over{\pd u}} \eeq
with $s=\log x,\, z=x^2/t$  and
$u=z\exp(-2s)\,  v(s,z)$ we get 

\bigskip
$\begin{array}{l}  
\!\! z^2(16 z^2 v_{z z z z}+4 z v_z^2+2 v   + 12  v_{z z} +
4 v z  v_{z z} +z^2 v_{z z} +
     48  z  v_{z z z}  +     2 v   v_z + \\ 
\!\! 4 z  v_z) - 6 v_s - v z v_s +z v_s^2 + 4 z^2 v_s v_z + 
       11 v_{s s} + z v v_{s s} +4 z v_{s z} + 4  z^2v v_{s z} -
            \\   
\!\! 6 v_{s s s} - 12 z v_{s s z} +24 z^2 v_{s z z}  +
               v_{s s s s} + 8 z v_{s s s z} +
               24 z^2 v_{s s z z} + 32 z^3 v_{s z z z}\= 0    
\end{array} 
$

\sk\ni
which indeed does not depend explicitly on $s$.
The dilational-invariant \so s are found putting
$v=w(z)$, and  only the terms in parenthesis survive.

For what concerns ``true'' CS, writing the general vector field in
the form
\beq X\, =\, \xi(x,t,u){\pd\over {\pd x}}+\tau(x,t,u){\pd \over{\pd
t}}+\phi(x,t,u){\pd \over{\pd u}} \lb{gsy}
\eeq
a complete list of CS has been given both for the case 
$\tau\not=0$ (and therefore, without any restriction, 
$\tau=1$)  \cite{LW} and for the case $\tau=0$
\cite{Cl,Lo}, see also \cite{OCRC}; it has been also shown that the invariant \so s
under these CS are precisely those found  
by means of the ``direct method" \cite{LW,Cl,CK,Lo}.

Let us give for completeness the form taken by the \BE\ when rewritten in 
terms of the canonical variables determined by some of these CS.
For instance, choosing the CS
\beq X\=t{\pd\ov \pd x}+ {\pd \ov\pd t}-2t{\pd\ov \pd u} \lb{BEC1}\eeq
we get  $s=t,z=x-t^2/2$ and $u=v(s,z)-s^2$, and
the \eq\ becomes
\beq
         -2  - v_z + v_z^2 +
         v v_{zz} + v_{zzzz} +v_{ss} - 2 s v_{sz} \= 0 \lb{BECS1}
\eeq
Starting instead (cf. Remark 2) from 
$(1/t)X=\pd/\pd x+(1/t)\pd /\pd t-2\pd/\pd u$, we get
$s=x,z=x-t^2/2,\, u=v(s,z)+2(z-s)$ and
\begin{eqnarray} \lb{BECS2}
-2  - v_z  + v_z^2 + v v_{z z}+  v_{zzzz}+2 v_s v_z+
               v v_{s s} - 2 s v_{ss} + 2 z v_{s s} &+& \\ \nn
               2 v v_{s z} - 4 s v_{sz} + 4 z v_{s z} +
               v_s^2  + v_{sss s} + 4 v_{s s s z} +
               6 v_{ssz z} + 4 v_{szzz} &=&0
\end{eqnarray}
Both \eq s (\ref{BECS1}) and (\ref{BECS2}) have the expected form
(\ref{DRT}),  and become the known reduced ODE (cf. \cite{LW})
if one looks for \so s with $v=w(z)$ and $u=w(z)-t^2$.

In the case of CS with $\tau=0$, the invariant \so s have the form
(instead of (\ref{uwz}))
\beq u(x,t)\=\a(x,t)+\be(x,t)\, w(t) \lb{wt} \eeq
where $w(t)$ depends only on $t$ and satisfies an ODE. Choosing, e.g., 
(cf. \cite{GMR})
\beq \lb{X2}
X\={\pd\over{\pd x}}+\Big({2u\ov{x}}+\de{48\ov{x^3}}\Big)
{\pd\over{\pd u}}\eeq
where $\de$ may assume  the values $0$ or $1$, 
the canonical variables are given  by $s=x,z=t$, and  
$\, u(x,t)=-12\de/x^2+x^2v(x,t)$,
and the \eq\ becomes 
\[       x^2 (v_{t t} +  6 v^2 ) + 
              8   x^3 v v_x + x^4 v_x^2 + 
              12 v_{x x} - 12 \de v_{x x} + 
              x^4 v v_{xx} + 8 xv_{xxx} + x^2 v_{xxxx} = 0
\]
which has the form (\ref{DRT}), observing that the role of the
variable $s$ is played here by $x$; as expected,
looking for \so s in which $v=w(t)$, this \eq\ becomes 
one of the \so s listed in \cite{Lo}.

To complete the analysis, one can also look for  \so s  of the form
\beq \lb{vx} u(x,t)\=\a(x,t)+\be(x,t)\,w(x) \eeq
with $w(x)$ satisfying an ODE, or for CS of the form
\beq \lb{Xt} X\={\pd\ov{\pd t}}+\phi(x,t,u){\pd\ov{\pd u}}\eeq
i.e. with $\xi=0$. It is not difficult to verify   that
no   true CS of this form is admitted by the \BE .  However, \so s of
the  form (\ref{vx}) can be obtained via {\em weak} CS. Indeeed, choosing
e.g.
\beq\lb{X1t} X\={\pd\ov{\pd t}}+\Big({1\ov{t^2}}-{2u\ov{t}}\Big){\pd
\ov{\pd u}}\eeq
one now obtains  $s=t,z=x$ and  $u(x,t)=1/t+v(x,t)/t^2$,
giving 
\beq v v_{x x} + v_x^2 +6v + t(v_{xx}+2) + t^2 v_{xxxx}- 4 t v_t
   + t^2 v_{tt}  \= 0
\lb{BExt}\eeq
which is precisely of the form (\ref{Kr}), showing that (\ref{X1t}) is a 
weak CS (the role of $s$ is played here by $t$).
Looking indeed for \so s with $v=w(x)$, one gets (cf. (\ref{d1s})) a 
system of the three ODE's 
\[v v_{x x} + v_x^2 +6v=0,\q 2+v_{xx}=0,\q 
v_{xxxx}=0 \]
admitting the common \so\ $w=-x^2$ and giving the 
\so\ $u=1/t-x^2/t^2$ of the \BE . 

Another example of weak CS is the following
\beq \lb{CSP} X\= t^2{\pd\ov \pd x}+{\pd\ov \pd t}-\Big(2x+{10\ov
3}t^3\Big){\pd \ov \pd u}\eeq 
now $s=t,z=x-t^3/3$ and $u=-2sz-s^4+v(s,z)$.
Instead of giving the form of the \eq\
in  these variables, let us now evaluate, according to our 
discussion (cf. (\ref{dddd})), the additional \eq s $\De^{(1)}=X^*(\De)=0$ 
etc.: we get
\beq \De^{(1)}\= -10  t - 3 u_x - 2 t u_{xt} - 
              {5\ov 3} t^3 u_{x  x} - x u_{x  x} \= 0 \lb{CP1}\eeq
\beq \De^{(2)}\= 
2 + u_{x t} + t^2 u_{x x} \= 0 \lb{CP2}\eeq
The most general \so\ of the \eq\ $\De^{(2)}=0$ is
\[ u\= F(t)+G\Big(x-{{t^3\over 3}}\Big)- 2tx \]
where $F,G$ are arbitrary functions of the indicated arguments; we easily 
conclude that (\ref{CSP}) is a weak CS of order $\s=3$ and,
taking into account also the invariance condition $X_Qu=0$, we 
obtain the invariant \so s 
\[ u(x,t)\= -{t^4\ov 3} - 2t x - {12\ov (x - t^3/3)^2}\q {\rm and}\q 
u(x,t)\= -{t^4\ov{3}}-2tx+c\lb{CP3}\]
($c=$ const). If instead we do {\em not} impose the 
invariance condition
$X_Qu=0$ and solve -- according to Remark 3 -- the three \eq s 
(\ref{BE},\ref{CP1},\ref{CP2}), we find the slightly more general 
families of \so s
\[ u(x,t)\= - {t^4\ov 3} -2tx+ c_1 t -{12\ov{(x-t^3/3-c_1)^2}} \q 
{\rm and}\q
u(x,t)\= - {t^4\ov{3}}-2tx+c_2t+c  \]
which are transformed by the \sy\ into one another,
showing that (\ref{CSP}) is also a partial \sy\ for the Boussinesq \eq .

\section{Concluding remarks}

Some of the facts presented in this paper were certainly already known, 
although largely dispersed in the literature,  
often in different forms and with different
languages. This paper is an attempt to provide a unifying scheme
where various notions and peculiarities of \sys\ of differential \eq s can be 
stated  in a natural and simple way. This allows us, in
particular, to provide a precise  characterization and a clear 
distinction between different notions of conditional \sy : this is indeed 
one of the main objectives of our paper. We can also give a 
neat comparison between the notions of conditional, partial and exact  \sys ;
several new and explicit examples elucidate the discussion.

In the same unifying spirit, it can be also remarked that all the above 
notions can be
viewed as particular cases of a unique comprehensive idea, which can be 
traced back to the general idea of appending suitable additional \eq s
to  the given differential problem $\De=0$, {\it and} to search for
(exact) \sys\ of this augmented problem (cf. \cite{FQ}).  In other words,
one looks for a supplementary \eq , say $E=0$, and a vector field $X$
satisfying
\beq X^*(\De)=G\De+HE\qq ;\qq X^*(E)=G_E\De+H_EE\lb{DE}\eeq
(some authors call generically ``conditional \sys '' for the \eq\ $\De=0$
all these \sys , and call ``Q-conditional'' \sys\ the more commonly named
conditional \sys .) Now, it is clear that all our above notions of
\sys\ simply correspond to different choices of the supplementary \eq\
$E=0$. Indeed:

\sk\ni
i) if $E=0$ is chosen to be  $X_Qu=0$, we are in the case of true CS,

\ni
ii) if $E=0$ is given by the system $X^*(\De)=X^*(\De^{(1)})=\ldots=0$, 
we are in the case of partial \sys , 

\ni
iii) if $E=0$ is the same as in ii) plus the condition $X_Qu=0$, we
are precisely in the case of weak CS.


\end{document}